\documentclass[twoside]{article}
\usepackage{fleqn,espcrc2}

\usepackage{graphicx}

\newcommand{\AmS}{{\protect\the\textfont2
  A\kern-.1667em\lower.5ex\hbox{M}\kern-.125emS}}
\newcommand{\beq}{\begin{equation}}
\newcommand{\eeq}{\end{equation}}

\title{Obstructions to dimensional reduction in hot QCD}

\author{Sourendu Gupta\address[MCSD]{Department of Theoretical Physics,\\
        Tata Institute of Fundamental Research,\\ 
        Homi Bhabha Road, Mumbai 400005, India.\\
        E-mail: sgupta@tifr.res.in}}%
       
\begin{document}

\begin{abstract}
\vspace{1pc}
I describe results on screening masses in hot gauge theories.  Wilsonian
effective long distance theories called dimensionally reduced (DR)
theories describe very well the longest screening length in pure gauge
theories. In the presence of fermions, meson-like screening lengths
dominate the long-distance physics for $3T_c/2\le T<3T_c$, and thus
obstruct perturbative DR.  Extrapolation of our results indicates that
a form of this obstruction may remain till temperatures of $10T_c$ or
higher, and therefore affect the entire range of temperature expected
to be reached even at the Large Hadron Collider.
\end{abstract}

\maketitle

\section{Dimensional reduction}

Finite temperature field theory in its Euclidean formulation exists on
an infinite spatial volume but a finite (Euclidean) temporal extent of
$1/T$. As a result, the Fourier modes of gluon fields have a countably
infinite set of momentum components in this direction--- $k_0=2\pi nT$
for $n=0,\pm1,\pm2,\cdots$. One might be able to integrate over the
non-zero modes to find a Wilsonian effective theory at at distances larger
than $1/T$ \cite{history}. The matching of correlation functions in the
two theories must be performed at a momentum scale $\Lambda_T\approx
{\cal O}(T)$ \cite{braaten}.  Consistency in applying perturbation
theory then demands that $\alpha_S(\Lambda_T)\ll1$. After matching
correlation functions in the two theories, the low-energy effective
theory must have exactly the same physics as the full 4-d theory--- the
same correlation functions, the same long distance screening behaviour,
and so on \cite{reisz}.

This procedure works extremely well in scalar $\phi^4$ theories
\cite{braaten}. It works for the gauge-Higgs system that describes
the dynamics of the electro-weak theory near its finite temperature
phase transition \cite{kajantie}. It has also been tested for $SU(2)$
pure gauge theory at $T\ge2T_c$.  The spectrum of screening masses in
$SU(2)$ has been extracted from correlation functions of local operators
built from gluon fields \cite{saumen}.  The perturbatively determined
dimensionally reduced (DR) theory has also been simulated numerically
and its screening masses have been determined \cite{hart}.  The lowest
screening mass (longest screening length) belongs to the thermal scalar
sector (see \cite{arnold} for the group theory).  At $2T_c$ we find
\beq
   \frac{\mu(0_+)}T =
       \cases{2.9\pm0.2   & (4-d theory),\cr
              2.86\pm0.03 & (DR theory).}
\eeq
Thus, at length scales of about $1/3T$ and more, the DR theory gives a
good description of the physics of the 4-d theory.

\section{Limits of DR and puzzles}

\begin{figure}[htb]
\includegraphics[scale=0.55]{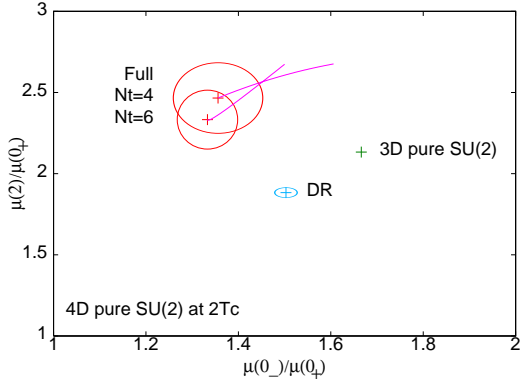}
\caption{Ratios of screening masses in the $SU(2)$ pure gauge theory
   at $2T_c$ determined with lattice spacings, $a$, of $1/8T_c$ ($N_t=4$)
   and $1/12T_c$ ($N_t=6$), compared with those obtained in the DR theory.
   The ellipses show 1--$\sigma$ error boundaries. The central points
   show the infinite spatial volume extrapolations--- the pink lines
   show the track of results on finite volumes.}
\label{fg.fig1}
\end{figure}

How does DR fare at distances less than $1/\mu(0_+)$? To answer this
we look at the next higher screening masses in the two theories. Some
notation will be useful--- screening masses describe propagation
from one two-dimensional slice of space to another \cite{arnold}. As
a result, they are labelled, not by the angular momentum $J$ of 3-d,
but by its two dimensional analogue, $J_z$, the projection along the
normal to the 2-d slices of space.  For every even $J$, the $J_z=0$
states are the thermal scalar $0_+$, and the $J_z=0$ of odd $J$ are
the thermal quasi-scalar $0_-$. The $J_z=\pm1$ states make the real
two-dimensional irrep, the thermal $\bf 1$. Similarly the $J_z=\pm2$
make up the thermal $\bf 2$. Measurements \cite{saumen,hart} showed that
$\mu(0_+)<\mu(0_-)<\mu({\bf 1}) \approx\mu({\bf 2})$. So, the question
is about the comparison of the higher screening masses in the 4-d and
DR theories.

In Figure \ref{fg.fig1} a comparison of the 4-d and DR theories is
shown. Notice the following points---
\begin{enumerate}
\item In the 4-d simulations, finite volume effects are under good
   control. The finite volume measurements extrapolate smoothly to the
   infinite volume limit.
\item Lattice spacing effects are under reasonable control, since the
   error ellipses for lattice spacings of $1/8T_c$ and $1/12T_c$ overlap
   significantly.
\item The 4-D theory lies many standard deviations away from the DR
   theory by this measure.
\end{enumerate}
In the light of the previous discussion, this mismatch is not unexpected,
since the 4-d and DR theories are matched at $\Lambda_T=2\pi T$, and the
higher screening masses all lie rather close to this scale. However, we
have gained a quantitative bound to the length scale at which DR fails
in the pure gauge theory--- DR cannot describe physics at length scales
of $1/\mu(0_-)$ or shorter.

This, in fact, is a puzzle. Recall that we have labelled states by the
dimensionally reduced symmetry group. The very fact that this symmetry
is obeyed even by the higher states shows that some version of DR occurs
in the theory, although it cannot be obtained by perturbative matching.

\section{Including fermions}

Even if the weaker form of perturbative DR works, then inclusion of
Fermions involves no new problems. Fermion modes are antisymmetric in
the Euclidean time direction and hence have $k_0 = \pi(2n+1)T$. There are
no zero modes and Fermions can be entirely integrated out. As a result,
there are no Fermion remnants in the long-distance theory--- every
correlation function involving Fermion field operators is integrated out
of the DR theory. The only traces of Fermions are subtle: they influence
the dimensionful couplings in the long-distance effective theory.
A simulation of QCD with four flavours of dynamical fermions shows that
this picture breaks down completely at $T\approx$ 1--3 $T_c$ \cite{rajiv}.

\begin{figure}[htb]
\includegraphics[scale=0.55]{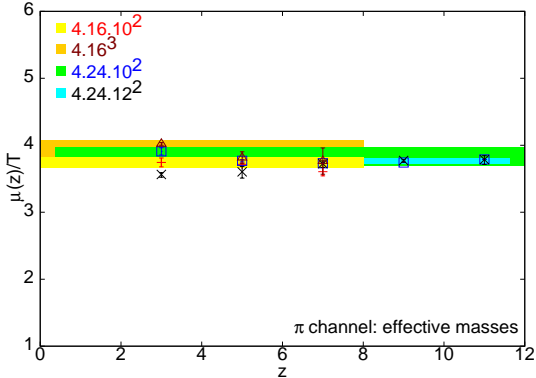}
\caption{The $\pi$-like $0_+$ screening mass at $2T_c$ on various
   lattice sizes. The effective masses are shown by various symbols
   and the bands show the results of fits to the correlation function
   (the width of the band is the 1--$\sigma$ error).}
\label{fg.fig2}
\end{figure}

We have simulated QCD on lattices of various sizes at temperatures of
$3T_c/2$, $2T_c$ and $3T_c$ and collected statistics over 1000--2000
mutually uncorrelated configurations. Simulation details can be
found elsewhere \cite{rajiv}.  Over these configurations of thermal
gauge fields, we have constructed correlators from quark-anti-quark
operators. The corresponding screening masses we call meson-like. We
have used operators corresponding to the $T=0$ $\pi$ and $\sigma$
and the vector and pseudo-vector mesons. All of these give non-trivial
correlations in the $0_+$ state. The $\pi$ and $\sigma$-like screening
masses are equal within our statistics as are the vector and pseudo-vector
$0_+$ screening masses. However, these two sets of $0_+$ screening
masses are not equal to each other.  The $\pi$-like screening mass,
$\mu_\pi(0_+)$ is smaller. Measurements on various lattice sizes (Figure
\ref{fg.fig2}) show the lack of finite volume effects.

\begin{figure}[htb]
\includegraphics[scale=0.55]{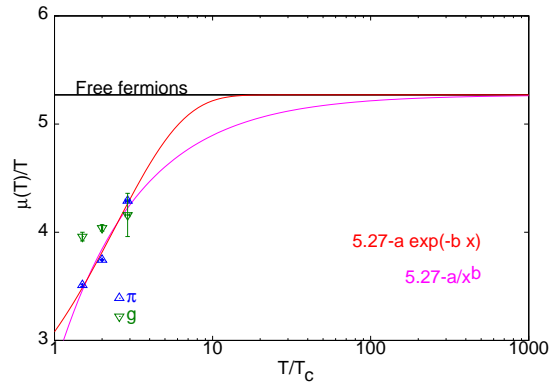}
\caption{Comparison of $0_+$ screening masses from $\pi$-like and gluonic
   correlators. The former are smaller until about $3T_c$, but lie
   significantly below the perturbative value until $10T_c$ or higher.}
\label{fg.fig3}
\end{figure}

We also measured the $0_+$ screening mass in the glue sector of the
theory, $\mu_g(0_+)$, and found that $\mu_\pi(0_+)<\mu_g(0_+)$ for
$T<3T_c$ (Figure \ref{fg.fig3}). This is a direct obstacle to DR at
$T<3T_c$: since the DR theory does not admit any Fermion operators,
it is ineffective in describing the true long-distance physics of the
QCD plasma, which arises in the Fermion sector.

Actually the Fermionic roadblock to DR is even wider. In a free-Fermion
theory, the meson-like screening correlators would decay at long
distances with an effective mass of $2\pi T$. ${\cal O}(a^2)$ lattice
artifacts would change this numerical value at the lattice spacing, $a$,
that we work at into the lower number indicated in Figure \ref{fg.fig3}
\cite{born}. This is the value $\mu_\pi(0_+)$ must have if perturbation
theory were to be reliable at the scale of $\mu_\pi(0_+)$.

Why do we want perturbation theory to work at the scale of $\mu_\pi(0_+)$?
Simply because we must set $\Lambda_T<\mu_\pi(0_+)$ to integrate out the
meson-like screening mass. If we are to use perturbative methods for
this, then we must expect that the expansion in $\alpha_s(\Lambda_T)$
captures the physics at this scale completely.  In particular, it must
also be able to reproduce the deviation of $\mu_\pi(0_+)$ from its
free-field value.  This cannot be done consistently within perturbation
theory unless $\mu_\pi(0_+)\approx2\pi T$. Hence, this less obvious roadblock
to DR.

We have tried various 2 parameter fits to our three data points to see
at what $T$ we get $\mu_\pi(0_+)\approx2\pi T$. The envelope of these fits
are given by the two curves shown in Figure \ref{fg.fig3}. This purely
phenomenological approach tells us that the obstruction to DR would
persist up to $T\approx10T_c$ or greater. It would, of course, be best
to simulate the theory at such temperatures. However, to avoid finite
volume effects with $N_t=4$ one would then have to take spatial sizes
greater than $40^3$ lattice units. Such a computation with dynamical
fermions is prohibitively difficult at present.

\section{Other phenomena}

\begin{figure}[htb]
\includegraphics[scale=0.55]{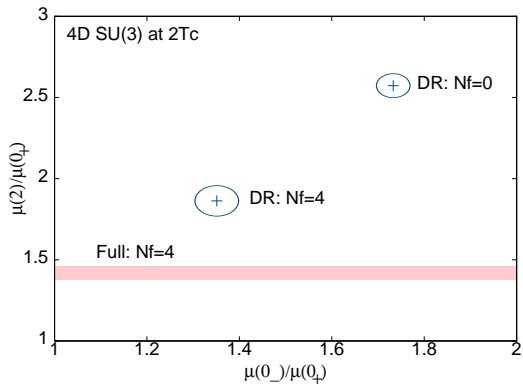}
\caption{Ratios of screening masses in 4-flavour QCD.}
\label{fg.fig4}
\end{figure}

Recently, screening masses have been computed in a DR theory obtained by
perturbative matching to QCD with dynamical Fermions \cite{laine}. Knowing
of the problems with this DR theory, it is nevertheless interesting
to ask what it predicts for the screening masses in the glue sector,
and how it compares with data from the 4-d theory. The temperatures in
\cite{laine} have been specified in units of $\Lambda_{\overline{\rm MS}}$
whereas those in \cite{rajiv} are in units of $T_c$. The quantitative
connection between these units is made possible by a recent determination
\cite{gupta} of the ratio $T_c/\Lambda_{\overline{\rm MS}}=1.07\pm0.05$
for 4-flavour QCD at the quark masses used in \cite{rajiv}.
At $T=2\Lambda_{\overline{\rm MS}}$ we find
\beq
   \frac{\mu_g(0_+)}T =
       \cases{4.04\pm0.05 & (4-d theory),\cr
              4.87\pm0.07 & (DR theory).}
\eeq
There is a statistically significant mismatch between these two
numbers, as we might now expect. The number quoted above for the 4-d
theory is at $T=2T_c=(2.14\pm0.10)\Lambda_{\overline{\rm MS}}$. Moving
to $T=2\Lambda_{\overline{\rm MS}}$ would mean lowering the temperature.
As seen in Figure \ref{fg.fig3}, this would lower $\mu_g(0_+)$
slightly, making the discrepancy slightly worse.

The test can be pushed further in terms of the ratios of screening masses,
as shown in Figure \ref{fg.fig4}. In the DR theory all the masses have
been extracted. In the 4-d theory a preliminary measurement of the glue
sector $\mu({\bf 2})$ has been made, and the ratio $\mu({\bf 2})/\mu(0_+)$
is shown as the horizontal band. The mismatch between the DR and 4-d
theories is obvious.

\section{Conclusions}

Dimensional reduction with perturbative matching of the DR theory to the
full 4-d finite temperature theory seems to work extremely well in many
cases. For 4-d $SU(2)$ pure gauge theory it works superbly at the scale of
the longest screening length, but fails at the scale of the next screening
length. This is expected if we consider DR as a Wilsonian effective theory
at the scale of the longest screening length. However, for QCD with
Fermions at $T\le3T_c$ it fails in all respects--- it does not contain
the Fermion composite states which give the longest correlation length,
it gives wrong results for the longest gluon screening length, and it
does not correctly reproduce the ratios of gluon screening lengths.
At higher temperatures, upto about $10T_c$, arguments given here lead
us to believe that it would be impossible to construct this effective
theory in QCD through a perturbative matching of correlation functions.

\end{document}